\RequirePackage{fix-cm}
\documentclass[10pt,a4paper]{article}
\usepackage[round,authoryear]{natbib}
\usepackage{graphicx}
\usepackage[]{subfigure,caption}
\usepackage{authblk}
\usepackage[margin=1in]{geometry}
\usepackage{helvet}
\begin{document}

\title{Flow in fractured porous media: A review of conceptual models and discretization approaches
}
%


\author[1,3]{Inga Berre}
\author[2]{Florian Doster}
\author[1]{Eirik Keilegavlen}

\affil[1]{Department of Mathematics, University of Bergen, Norway}
\affil[2]{Institute of Petroleum Engineering, Heriot-Watt University, UK} 
\affil[3]{Christian Michelsen Research AS, Norway}

\date{\today}

\maketitle

\begin{abstract}
The last decade has seen a strong increase of research into flows in fractured porous media, mainly related to subsurface processes, but also in materials science and biological applications.
Connected fractures totally dominate flow-patterns, and their representation is therefore a critical part in model design.
Due to the fracture's characteristics as approximately planar discontinuities with an extreme size to width ratio, they challenge standard macroscale mathematical and numerical modeling of flow based on averaging. Thus, over the last decades, various, and also fundamentally different, approaches have been developed. 
This paper reviews common conceptual models and discretization approaches for flow in fractured porous media, with an emphasis on the dominating effects the fractures have on flow processes. In this context, the paper discuss the tight connection between physical and mathematical modeling and simulation approaches.  Extensions and research challenges related to transport, multi-phase flow and fluid-solid interaction are also commented on.
\end{abstract}
\section{Introduction}
\label{intro}
Fractures, both natural and engineered, provide major conduits or barriers for fluid flows in many types of media. 
The existence of fractures that affect flow and transport, are characteristic of different types of porous media for domains ranging from millimeters to hundreds of kilometers. Subsurface rocks are, for example, always to some degree fractured due to tectonic movement. Open fractures act as conduits and sedimented fractures as barriers to flow and severely affects processes. Fractures are present porous media like soil \citep{National2001} and glaciers \citep{Fountain1998} and in porous materials such as wood \cite{smith2003,Watanabe1998}.
Several anthropogenic materials, like concrete \cite{carmeliet2004three, Rouchier2012} have similar characteristics, intendedly constructed or not. Thin membranes  or other thin layers in porous materials, for example in fuel cells \cite{Qin2014,Qin2015} or biological materials,  can conceptually also be modelled in the same manner. 

For a porous medium, secondary permeability provided by conductive fractures can result in an effective permeability several magnitudes larger than the primary permeability provided by connected pores. 
Moreover, there is generally no separation of length scales in fracture sizes, thus upscaling into average properties often results in poor model quality. 
As a supplement, a wide range of methodologies tailored for fractured media have been proposed in recent years. These have in common that they aim to accurately capture the effect of complex fracture networks on the flow pattern. Aided by increases in computational power that make higher resolution models feasible, developments typically focus on one or more of the following aspects: (1) accurate representation of flow in the fractures, (2) accurate representation of flow in the interaction between fractures and the surrounding porous media, and (3) ability to handle complex fracture network structures.  

This paper reviews the most important modeling approaches for flow in fractured porous media, from physical, conceptual and mathematical models, to  discretization approaches.  We provide an integrated view on modeling and discretization, and therefore do not give details on specific numerical schemes, but rather present broader classes of methods, covering strengths and weaknesses in terms of modeling capabilities, computational accuracy and cost. Approaches to be covered include methods that are designed to work with traditional simulators for flow and transport, as well as models that are based on explicit representation of fractures. This gives rise to discretization approaches with different characteristics and features, depending on how fractures are represented in the computational domain, and how flow between fractures and between fractures and surrounding porous media is approximated. 

The paper provides a broad introduction for researchers that are new to the topic of modeling and simulation of flow in fractured porous media, with emphasis on dealing with challenges arising from the strong effect networks of fractures have on aggregated flow patterns. The general overview of physical, mathematical and conceptual models and discretization approaches aim to contribute in bridging research developments across different fields and application areas. The paper is not intended as a comprehensive review of the topic, and for the sake of brevity, the references are not exhaustive. More detailed discussion of some of the aspects addressed herein and related topics can be found in previous books
\citep{Adler2012,Dietrich2005, Sahimi2011, Faybishenko, Bear2012}  and review articles \citep{Berkowitz2002,Neuman2005}. 

The paper starts by discussing the representation of fractures and fracture networks in porous media in Sec. \ref{sec: Fractures in porous media}, while models for flow are presented in Sec. \ref{sec: Modeling of flow in fractures and matrix}. In Sec. \ref{sec:conceptual models}, conceptual models for representing porous media with complex networks of fractures  are reviewed. Discretization approaches are discussed in Sec. \ref{sec:discretization}. In Sec. \ref{sec:discussion}, the choice between different modeling approaches are discussed as well as research challenges that remain, before the paper is concluded in Sec. \ref{sec:conclusion}.


\section{Geometric representation fractured porous media}
\label{sec: Fractures in porous media}
Fractures in a porous medium are discontinuities in the medium in the form of narrow zones that are locally approximately planar and have distinctly different characteristics than the medium itself. 

The detailed geometry of a fracture is represented by its boundaries to the medium on each of its sides. While fractures are generally thin compared to their extension, they are much wider than a typical pore diameter, and their length can span orders of magnitude, with a largest extent that can be comparable to the size of the domain of interest. 
At the same time, the total volume corresponding to the fractures is small compared to the volume of the surrounding medium.  Due to these characteristics, a fracture in a porous medium is typically approximated by a two-dimensional inclusion in macroscale modeling of flow as illustrated in Fig. \ref{fig:NetworkNotationSimulation}a.

To the utmost, a  fracture can be empty or it can contain an impermeable filling material.  As a consequence, fractures can be both highly conducting or highly sealing compared to the host material. Networks of  multiple, connected fractures  structurally dominate 
flow patterns based on network geometries that can be highly complex; we illustrate this in see Fig. \ref{fig:NetworkNotationSimulation}c, by a simulation using the software PorePy \cite{keilegavlen2017porepy}.

To model flow in porous media with complex networks of fractures calls for adequate representations.  A major challenge is the difficulty of defining and measuring model parameters that allows for continuum-scale modeling, both in situations where fractures are treated explicitly and when they are integrated as part of the porous medium itself. 
In modeling of standard porous media, parameters used in macroscale modeling are related to pore-scale quantities by averaging over domains of increasing size. 
If, with increasing size of the averaging domain, the averaged quantity becomes approximately constant, the corresponding domain is denoted a \textit{representative elementary volume} (REV) \citep{Bear1972}. Parameters defined by the REV concept allow for building of flow models with quantities that varies continuously in space. 
Compared to the standard descriptions of a porous medium, fractures introduce intermediate length scales, but there may be no clear scale separation between the pore scale, fracture widths, fracture lengths and the macroscale of interest. Hence, the presence of fractures challenges the existence of a REV in porous-media modeling, and the resulting difficulty in defining averaged quantities will be a recurring topic in this paper. 
In the following, a brief discussion of the representation of individual fractures and fracture networks in a fractured porous media is given. 

\subsection{Individual fractures}
A fracture is bounded by two surfaces, which may be rough and locally oscillating. A thorough description of the geometrical and statistical modeling of a single fracture is given by \citet{Adler2012}. Herein, we do not focus on the representation of the detailed geometry of a single fracture, but consider the fracture at a scale much larger than the distance between the fracture's bounding surfaces and a representation of the fracture at this scale that is suitable for modeling of flow.

If a two-dimensional coordinate system is defined parallel to the plane defined by the 
fracture, the pointwise distance between the two bounding surfaces of the fracture defines its \emph{local aperture}. If the fracture's bounding surfaces are locally fluctuating, the local aperture varies significantly over short distances. By representing the fracture's bounding surfaces by average planes, the \emph{average aperture}, $e_V$, of the fracture can be defined as the distance between these two planes. This gives an idealized measure of fracture width that does not account for the possible fine-scale fluctuations of the fracture's bounding surfaces. 

In modeling of different physical processes, such as fluid flow and mechanical deformation of a fracture,  different concepts for defining an idealized parameter that represents the width of the fracture in the context of the given process are commonly employed. This results in definitions of aperture that are conceptually different from the average aperture defined above, and take different numerical values. An example will be given in Sec. \ref{sec: Modeling of flow in fractures and matrix}, which introduces the \emph{hydraulic aperture} of a fracture.

\subsection{Fracture networks}
In networks of fractures, the size, orientation, location and aperture that characterize individual fractures are complemented by measures of the density of fractures. For fractured rocks, both properties of individual fractures and networks are modeled by various statistical distributions of (lognormal, gamma, exponential, power law); networks may also be characterized as fractal descriptions \citep{Bonnet2001}. 
An argument for power law and fractal scaling is the lack of characteristic length scales in fracture growth. In the case fractures exist on multiple scales, even if the most dominating fractures may be represented explicitly, as discussed in Sect. \ref{sec:conceptual models}, there may not exist an REV for the remaining part of the domain. 

The connectivity of a fracture network is a central defining property 
considering flow and transport processes, and can be studied by percolation 
theory \citep{Robinson1983,Berkowitz1995}. When a network is 
statistically homogeneous with well defined parameters, percolation thresholds 
can be studied and predicted. Close to percolation thresholds, connectivity 
will be highly sensitive to fracture density. Furthermore, for general networks 
of fractures that are not well described statistically, individual members may 
be decisive for the connectivity of the network as a whole.  

\section{Modeling of flow in individual fractures and the porous matrix}
\label{sec: Modeling of flow in fractures and matrix}
Fractured media provide some particular challenges to modeling and simulation that are not present in standard porous media. These can be explained by the interaction between the geometric properties of fractures and 
topological properties of fracture networks with dynamic processes that takes place in the domain: The fractures will strongly affect the nature of processes, for example fluid velocities. Leading processes may not be well described in a continuum manner, as the fractures by definition introduce strong discontinuities that are not well represented by averaged descriptions \citep{Long1982}. 
Moreover, flow processes can also affect the nature of the fracture networks, for example due to mechanical or chemical fluid-solid interactions.

\begin{figure}
\begin{tabular}{cc}
\begin{tabular}{c}
\begin{subfigure}[Fracture network]{
\includegraphics[width=.3\textwidth, trim=0 -0.5cm 0 0 ]{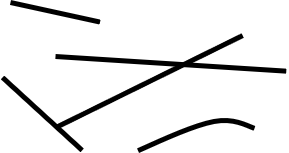}}
\end{subfigure}\\
\begin{subfigure}[Notation]{
\includegraphics[width=.4\textwidth, trim=0 -0.5cm 0 -0.5cm ]{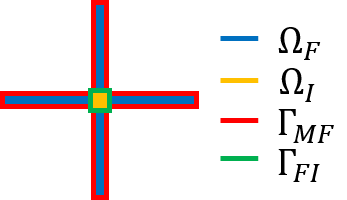}
}
\end{subfigure}
\end{tabular}
& 
\begin{tabular}{c}
\vspace{.85cm} \\
\begin{subfigure}[Simulation of heat transport]{
\includegraphics[width=.47\textwidth, trim=0 -0.9cm 0 0 ]{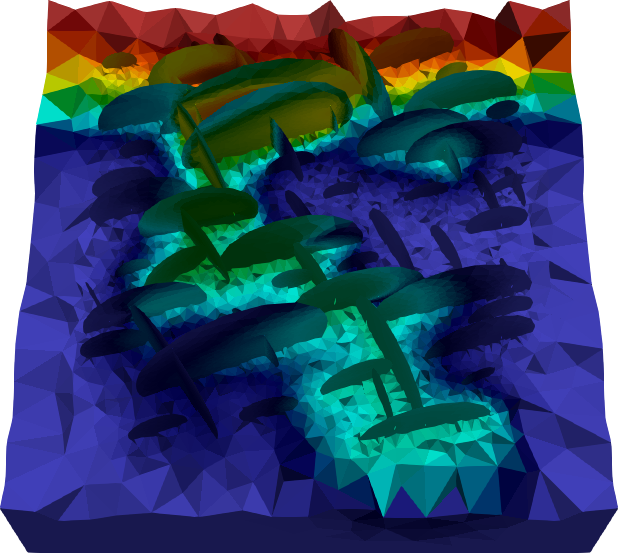}
}
\end{subfigure}
\end{tabular}
\end{tabular}
\caption{a) Conceptual drawing of a fracture network, including different intersections and non-planar fractures; b) Notation for fractures, intersections and interfaces; c) Simulation of heat transfer by convection, dominated by fracture flow.}
\label{fig:NetworkNotationSimulation}
\end{figure}

In this section, we present basic equations for modeling of flow in fractured porous media, considering a fractured porous medium domain $\Omega$. The modeling is based on a domain-decomposition approach, based on the different structural features of the fractured medium. 
Following standard conventions, we shall refer to the continuum host medium as the matrix, represented by $\Omega_M$, noting the ambiguity of the use of the term matrix as this term is also used for the solid skeleton when describing the microscale structure of a porous medium. The matrix, $\Omega_M$, can be porous, and it can also possibly integrate the effect of fine-scale fractures. The part of the domain $\Omega$ consisting of explicitly represented fractures is divided into fracture intersections,  $\Omega_I$, and the remaining part of the fractured domain  denoted  $\Omega_F$. The interface between  $\Omega_F$ and  $\Omega_M$  is denoted $\Gamma_{MF}$, while the  interface between $\Omega_I$ and $\Omega_F$ is denoted $\Gamma_{FI}$. A conceptual drawing of a fracture network is provided in Fig. \ref{fig:NetworkNotationSimulation}a, while the division of the fractured part of the domain and defined interfaces are illustrated in Fig. \ref{fig:NetworkNotationSimulation}b. 
We assume that flow only takes place between objects one dimension apart, that is, between $\Omega_M$ and $\Omega_F$ and between $\Omega_F$ and $\Omega_I$, but not directly between $\Omega_M$ and $\Omega_I$.

\subsection{Matrix flow}
\label{sec:matrix_flow}
Assuming that there exist a REV for the matrix part of the domain, single-phase flow at the associated continuum scale is governed by the mass conservation equation and Darcy's law. 
Mass conservation for single-phase flow in its continuum form is given by
\begin{equation} 
\label{eq:mass_conservation}
\frac{\partial (\phi \rho)}{\partial t}+\nabla \cdot {(\rho \bf v)}=q, \quad  \mathbf{x} \in \Omega_M, \end{equation}
where $\rho$ is fluid density, $\mathbf{v}$ is the flux and $q$ represents a source term.
In $\Omega_M$ Darcy's law relates the  flux $\mathbf{v}$ to the gradient of the pressure $p$ by
\begin{equation}
\label{eq:Darcy}
\mathbf{v}=-\frac{\mathbf{K}}{\mu}(\nabla p-\rho \mathbf{g}), \quad \mathbf{x} \in \Omega_M,
\end{equation}
where $\mathbf{K}$ is permeability, $\mu$ is viscosity and $\mathbf{g}$ is the gravitational acceleration.

\subsection{Flow in a single fracture}
A fracture in a porous medium can be empty, or contain filling material, which makes the fracture itself a porous medium. 
If the fracture is treated as a thin, but fully three-dimensional,  inclusions with a $3 \times 3$ permeability tensor, flow is governed by (\ref{eq:Darcy}) with a discontinuity in permeability associated with the fracture. However,  models that exploit the representation of a fracture as a lower-dimensional surface is more appropriate in particular when considering multiple fractures that spans a complex network in a porous medium.  

First, consider the case of laminar flow in the fracture residing in an impermeable matrix.  By averaging of the Stokes equations in the case of an empty fracture 
\citep{Adler2012}, or of Darcy's law in the case the fracture is filled with 
a porous material \citep{Martin2005,Angot2009}, the average specific discharge ( flux) can be expressed as    
\begin{equation}
\label{eq:Darcy_f}
\mathbf{v}^{\parallel}_F=-\frac{\mathbf{K}^\parallel}{\mu}(\nabla^\parallel p-\rho \mathbf{g}^\parallel), \quad \mathbf{x} \in \Omega_F,
\end{equation}
where $\mathbf{K}^\parallel$ is a $2 \times 2$ tensor and  $\mathbf{g}_\parallel$ is the component of $\mathbf{g}$ in the fracture plane. The pressure, $p$, should be interpreted as a macroscale pressure and $\nabla_\parallel$ is the in-plane gradient operator. Due to the resemblance of equation (\ref{eq:Darcy_f}) with Darcy's law,  Eq. (\ref{eq:Darcy}), $\mathbf{K}^\parallel$ is denoted the fracture permeability. 

The form of $\mathbf{K}^\parallel$ depends on the configuration of the fracture's bounding surfaces and on possible filling material. In the case of an idealized representation of a conductive fracture as a channel between two parallel planes separated by a constant distance $e_h$, and further assumptions of incompressible flow and no-slip conditions on the confining planes, the fracture permeability can be calculated directly to be isotropic; that is, $\mathbf{K}^\parallel=\kappa_f\mathbf{I}$. Its permeability is given by 
\begin{equation}
\label{eq:cubic_law}
\kappa_f=\frac{e_h^2}{12},
\end{equation}
where $e_h$ is denoted the hydraulic aperture of the fracture; see, e.g., \cite{Adler2012} for details. 

Equation (\ref{eq:cubic_law}) illustrates the strongly nonlinear relationship between fracture aperture and flow. If we consider flow across a 
cross-section of a fracture, we observe that the flow rate through a fracture 
per unit length is given by  $e_V\mathbf{v}_f\propto e_V\mathbf{K}^\parallel $. 
That is, in the case of parallel plate empty fracture, where $e_h=e_V$,  Eq. (\ref{eq:cubic_law}) results in a cubic relation between aperture and flow, and, hence,  Eq. (\ref{eq:cubic_law}) is often denoted the cubic law.
For more general conductive fractures, there IS a nonlinear dependence of the 
permeability on the aperture, and the aperture is thus a critical parameter. 
An important consideration in the construction of a simulation model is thus how accurately the aperture, including macroscale local variations along the fracture, and the associated fracture permeability, is represented. 

To accommodate realistic fractures, the fracture permeability must account for the roughness of fracture walls, and the presence of gauge or other filling material inside the fractures.  A result is that the hydraulic aperture $e_h$ corresponding to the fracture permeability may be much smaller than the average volumetric aperture $e_V$. To estimate the hydraulic aperture of a fracture is a non-trivial task in terms of both modeling and experiments \citep{Adler2012,Berkowitz2002}. Furthermore, for open fractures having anisotropic geometrical characteristics , the hydraulic aperture must be described by a tensor permeability $\mathbf{K}^\parallel$ defining preferential directions for flow. This is often the case for rock fractures due to shearing \cite{auradou2005permeability}, or for filled fractures with anisotropy in the filling material.

\subsection{Flow across fracture-matrix interfaces }
In the general situation, the matrix will be permeable, and there can be flow  between fracture and matrix. 
The mass conservation equation (\ref{eq:mass_conservation}) is then the same as before for the fracture, but with additional source terms representing flow into the fracture from the matrix; that is,
\begin{equation}
\label{eq:mass_conservation_f}
\frac{\partial (\phi \rho )}{\partial t}+\nabla^\parallel \cdot {(\rho \bf v_F^\parallel)}=q_F +\frac{\rho}{e_V}( v^\perp_{\Gamma^+_{MF}}+ v^\perp_{\Gamma^-_{MF}}) , \quad \mathbf{x}\in \Omega_F,
\end{equation}
where $v^\perp_{\Gamma_{MF}}$ denotes the flux across the interface $\Gamma_{MF}$, and the superscript on $\Gamma_{MF}$  denotes the restriction of $\Gamma_{MF}$ to the different sides of the fracture.

As the fracture is represented as a thin planar layer, flow in the direction tangential to the fracture is governed by equation (\ref{eq:Darcy_f}), but  an additional equation is needed to account for flow between fracture and matrix at each side of the fracture; that is, $v_{\Gamma^+_{MF}}^\perp$ and $v_{\Gamma^-_{MF}}^\perp$. Following \citet{Martin2005},  by averaging, fluxes across  the fracture-matrix interfaces at each side of the fracture are approximated by
\begin{equation}
\label{eq:interface_flux}
v_{\Gamma^{\pm}_{MF}}^\perp = \frac{\kappa^\perp}{\mu}\bigg(\frac{2(p_M - p_F)}{e_V}-\rho \mathbf{g}\cdot \mathbf{n}^{\pm} \bigg),
\end{equation}
where $\mathbf{n}^{\pm}$ denotes the normal to $\Gamma^{\pm}_{MF}$ pointing into of $\Omega_M$.  Furthermore, $\kappa^\perp$ is the permeability  in the normal direction which represents resistance to flow across the fracture, $p_M$ is the pressure in the matrix at the boundary of the fracture and $p_F$ is the fracture pressure. The factor 2 comes from taking the pressure difference over half the fracture aperture.
The matrix pressure will thus experience a jump across the fracture surface, with a magnitude that depends  strongly on whether the fracture is conductive or sealing.

\subsection{Flow across intersecting fractures}
\label{sec:flow_across_intersecting_fractures}
The adaption of Eq. (\ref{eq:Darcy_f}) for flow in an intersection between two fractures is 
\begin{equation}
\label{eq:Darcy_i}
v^{\mid}_F=-\frac{\kappa^\mid}{\mu}(\nabla^\mid p-\rho g^\mid), \quad \mathbf{x} \in \Omega_I,
\end{equation}
where $v_F^\mid$ is the flux along the one-dimensional interface, $\kappa^\mid$ is the permeability of the intersection, $\nabla^\mid$ is the one-dimensional gradient operator along the intersection and $g^\mid$ is the component of the gravity acceleration vector  along the intersection. For flow between a fracture and a fracture intersection, Eq. (\ref{eq:interface_flux}) can be adapted straightforwardly, considering the pressure difference between the boundary of the intersection and the pressure in the intersection \cite{Boon2016}. 

\subsection{Modeling of fractures as lower-dimensional objects}
\label{sec:mixed-dimensional}
Given the model presented earlier in this section, fractures can straightforwardly be treated as co-dimension one domains in the medium  based on a domain-decomposition approach for intersections, fractures and matrix \citep{Angot2009, Martin2005}. That is, for a three-dimensional domain, fractures can be considered as two-dimensional surfaces. Similarly, fracture intersections can be treated as co-dimension two domains (lines) in three dimensions  \citep{Boon2016}. This \emph{mixed-dimensional modeling} of domains is highly advantageous in modeling as it avoids thin equi-dimensional domains representing the fracture. 

In a co-dimension one representation of the fracture, aperture can be treated as a parameter  that possibly can vary, without the need to update the matrix domain.  This is advantageous in coupling flow with dynamical chemical or mechanical processes that change the fracture aperture \citep{Ucar2017}.

\subsection{Extensions to more complex flow physics}
The linear relation between flow and pressure gradient in equation (\ref{eq:Darcy_f}) relies on the assumption of laminar flow \citep{Whitaker1986}. 
If this assumption does not hold, because of high imposed pressure gradients relative to the resistance from fracture walls and gauge material, the result will be 
reduced flow rates compared to the linear relation in Eq. (\ref{eq:Darcy_f}). The standard extended model 
involves a Forchheimer correction term, which introduces a non-linear coupling 
between pressure gradient and flow rates.  Considering fractures as co-dimension one objects, the domain-decomposition framework presented above can be extended to include the Forchheimer correction in the fracture part of the domain \citep{Frih2008}.
Furthermore, considering the development of boundary layers developing at the boundary between the matrix and fracture domains, the effective fracture permeability is shown to depend on ratio of bulk and interface permeabilities, the fluid viscosity, and the fracture aperture \citep{Vernerey2012}.

\section{Conceptual models for representation of fractured porous media}
\label{sec:conceptual models}
\begin{figure}
\centering
	\includegraphics[width=1\textwidth]{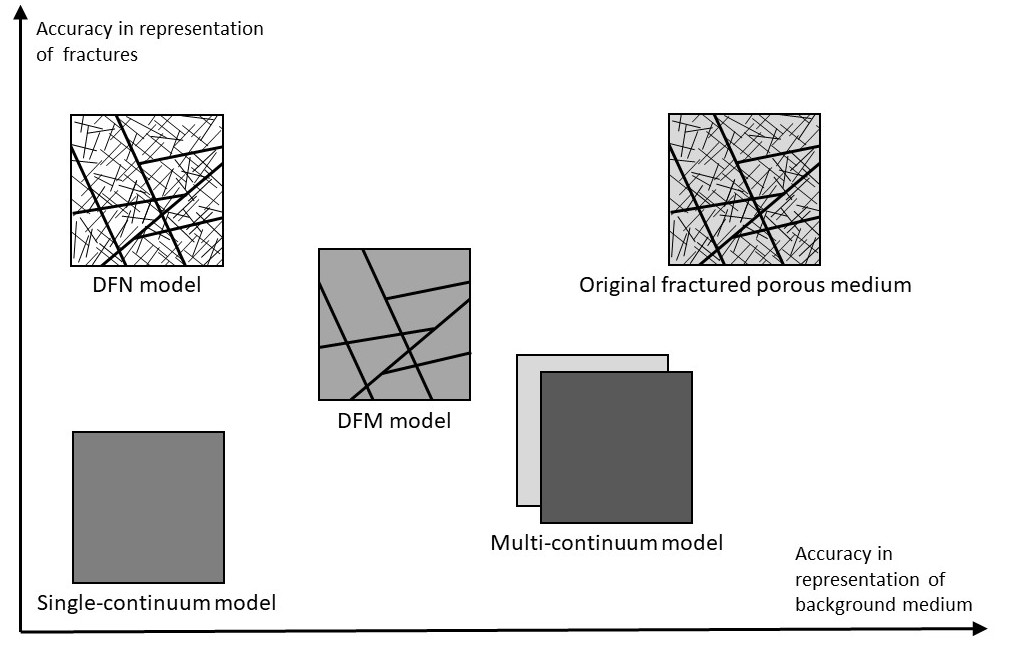}
	\caption{Conceptual models for modeling of a fractured porous medium. }
	\label{fig:conceptual_models}       
\end{figure}

The dynamics of flow within a single fracture and interaction with the surrounding matrix is in itself complex, but feasible to treat with the models in the previous section. Additional challenges arise when these dynamics take place in a network of fractures, where the fractures together have decisive impact on the dynamics in the simulation domain.
The fundamental choice in modeling and simulation of macroscale processes in fractured porous media is whether to represent fractures explicitly, or represent them implicitly by an effective continuum. This is not a binary choice, and several approaches are available that employ a combination of the two. 
In the following, different conceptual models, as illustrated in Fig. \ref{fig:conceptual_models}, are described.

\subsection{Implicit representation of fractures}
\subsubsection{Single-continuum models}
\label{sec:ImpRepSingCont}
In single-continuum models the fractures are represented by altering the permeability of the porous medium. The permeability may increase or decrease, and the orientation of the tensor may change depending on the properties of the fractures and the fracture network. Ultimately, one aims to identify an effective permeability $\mathbf{K_e}$ of the fracture network together with the porous matrix such that 
\begin{equation}
\langle\mathbf{v}\rangle=-\left\langle\frac{\mathbf{K}}{\mu}(\nabla p-\rho \mathbf{g})\right\rangle=-\frac{\mathbf{K_e}}{\mu}\left\langle(\nabla p-\rho \mathbf{g})\right\rangle, \label{eq:upscaled_darcy}
\end{equation}where the angular brackets $\langle \cdot\rangle$ denote the average over a suitable volume that contains fractures and porous matrix. In words, one expects that also at the larger scale the mean value of the velocity field is proportional to the mean value of the gradient of a potential. At this level the approach is identical to classical upscaling of heterogeneous porous media. 
The averaging volume should ideally be corresponding to an REV, which as discussed may not exist due to the lack of scale separation in the fracture network. An effective permeability can still be computed for a given averaging volume, say a grid block, and used in simulations, but as a modeling concept this approach is less powerful than when an REV is available. 
To emphasize the difference the term equivalent (grid-block) permeability \citep{Durlofsky1991} is used in the context of discretizations. 

An upscaling technique should account for inherent properties of the fractures themselves such as orientation, aperture, area, surface roughness and filling material as well as properties of the fracture ensemble such as fracture density and coordination number. 
Classical techniques for permeability upscaling, including methods based on numerical upscaling, asymptotic expansions and effective medium techniques, can all be applied to fractured media. Challenges pertinent to those techniques, as discussed in e.g. \citet{Farmer2002,Gerritsen2005} are amplified in fractured porous media.

\citet{Oda1985} devised an effective medium theory based method that calculates the effective permeability of the fracture network based on fracture shape, size, aperture and orientation distributions of the fractures, as well as matrix permeability. It does not account for the network properties but is very popular due to its simplicity. A recent review on explicit expressions including experimental observations as well is provided in \citet{Liu2016}. Even though it does not account for flow in the matrix, percolation theory has played a major role \citep{Sahimi1993}. Also, the fractal nature of fracture distributions has been used to develop upscaling techniques. Further, effective medium theory has been adapted to cope with challenges posed by   
individual fracture \citep{Saevik2013}. 

The single-continuum representation is very desirable because of its level of computational and model complexity. No additional terms in comparison with non-fractured porous media models are required. Hence, the transfer of methods and simulation technology is straight forward. However, it comes at the cost of simplification: For single-phase steady-state flow, it is always possible to calculate an equivalent permeability for a given boundary value problem such that the average flux follows the upscaled Darcy equation (\ref{eq:upscaled_darcy}). This equivalent permeability is, however, length scale, boundary condition and process specific. For more complex problems, even as simple as the evolution of pressure in compressible flow or tracer transport, the dynamics are not adequately represented by averaged flow alone. Thus, the boundary value problem to determine the equivalent permeability should be representative for the real flow field, and, as this may be hard to achieve \textit{a priori}, the results must be interpreted carefully. 

\subsubsection{Multi-continuum models}
Multi-continuum approaches are families of methods that represent the fractured porous medium by several superimposed media with their own conservation equations and constitutive laws.  
Thus, the mass conservation equation (\ref{eq:mass_conservation}) for continuum $\alpha$ reads as   
\begin{equation}\frac{\partial \left(\phi^\alpha \rho^\alpha\right)}{\partial t}+\nabla \cdot \left(\rho^\alpha{\bf  v^\alpha}\right)=q^\alpha + T^\alpha,
\end{equation}
where  $T^\alpha$ denotes the transfer into the continuum $\alpha$ from all other continua.
The flux ${\bf v^\alpha}$ represents the upscaled flux in continuum $\alpha$, which is almost exclusively described by Darcy's law (\ref{eq:Darcy}) with according upscaled permeabilities. 

The simplest approach consists of a fracture continuum $\alpha=f$ and a matrix continuum $\alpha=m$  and is referred to as dual-continuum model. Figure \ref{fig:conceptual_models} illustrates the concept. The concept can be extended to a general number of continua to account for heterogeneities in properties and flow rates within the fracture continuum.

The underlying assumption of the multi-continuum approach is that it is possible to represent different domains in the pore-space and the dynamics within, as well as the exchange between those domains based on continuous parameters and variables. Hence, the discussions and challenges for the single-continuum representation in Sec. \ref{sec:ImpRepSingCont} with respect to the existence of REVs apply to multi-continuum formulations as well. Due to the increased complexity of the equations the issue is even more involved in the multi-continuum setting, and derivations of the models from a fine scale is available only under very specific assumptions on geometries and flow \citep{Arbogast1990}. 
More generally, multi-continuum models should be considered constitutive formulations that can still be useful, in particular when some continua provide volume while others provide permeability. Note that sealing fractures provide neither, and, hence, are usually not modeled through multi-continuum approaches. 

The transfer term $T^\alpha$ is the art and the challenge of multi-continuum modeling. It shall represent the aggregated transfer behavior based on variables of the continuum formulation. In order to be predictive, its parameters shall be derived from geometrical, topological and physical properties of the individual continua.  The only physical constraint is mass conservation. For dual-continuum formulations this implies that  $T^m=-T^f$ holds true. The first explicit form for the transfer term was proposed by \citet{Warren1963} and \citet{Barenblatt1960} as
\begin{equation}\label{eq:transfer}
T^f=\sigma \beta (p^m-p^f),
\end{equation}
where the shape factor $\sigma$ accounts for geometrical properties such as effective fracture-matrix interface area and effective characteristic length over which the transfer takes place and $\beta$  accounts for physical properties such as matrix and fracture permeabilities and fluid viscosity.
For simple regular and symmetric structures of the fractured porous medium and under additional constraints on the magnitude and heterogeneity of parameters and variables, Eq. (\ref{eq:transfer}) accurately represents transfer between fracture and matrix, with $\sigma$ and $\beta$ taken as constants. For some of those configurations, explicit formulas to determine $\sigma$ and $\beta$  have been developed. However, in many applications pressure differences are substantial and fracture geometry is asymmetric and complex and hence, the representation of the transfer through Eq. (\ref{eq:transfer}) is inaccurate.

The common choice in extending the validity of Eq. (\ref{eq:transfer}) is to keep the linear pressure difference, and generalize the coefficient in front. No general approach is known, instead the available expressions are strongly tied to assumptions on the flow regime and fracture network configuration. 
For processes with negligible buoyancy, linear density-pressure relationships and vanishing macroscopic flux in the matrix,  upscaling techniques such as homogenization through asymptotic expansion \citep{Arbogast1990} 
have been used to determine the transfer term. These yield transfer terms with memory functions that are non-local in time and  therefore notoriously challenging to interpret or exploit in solutions and simulators.
To simplify calculations these have been expanded in series of exponentials yielding a series of first order transfer terms with varying transfer rate coefficients \citep{Tecklenburg2016}. Also heterogeneous fracture spacings yield series of first order transfer terms with varying coefficients due to varying shape factors \citep{Haggerty1995}.  
Both are referred to as multi-rate transfer models.  Another way of coping with the shortcomings of the transfer term in Eq. \ref{eq:transfer} is to expand the matrix continuum into a series of multiple interacting  continua (MINC) with vanishing macroscopic flow \citep{Pruess1985}.

For monotonic transfer processes analytic solutions for the diffusion equations are available that produce explicitly time dependent shape factors \cite{Rangel-German2006} or explicit parameterizations of the transfer with respect to time \citep{Lim1995,Zhou2017}.  While these approaches improve the accuracy significantly, their implementation in simulators tends to be inelegant. 

\subsection{Explicit representation of fractures}
Continuum models make no geometric distinction between $\Omega_M$ and $\Omega_F$, but instead employ different media that are coexisting in space. In contrast, conforming methods preserve a notion of $\Omega_M$, $\Omega_F$ and $\Gamma_{MF}$ as separate geometric objects.
Depending on the specific approach being employed, this reduces or removes altogether, the need to represent flow by upscaled quantities.
As such, modeling within a framework with explicit modeling  of fractures is often conceptually simpler than the implicit counterpart, but at the cost of dealing with complex geometries.

\subsubsection{Discrete fracture matrix models}
Discrete fracture matrix (DFM) models attempt to strike a balance between loss of accuracy by upscaling, and geometric complexity. The fluid is located in explicitly represented fractures, defined as $\Omega_F$, as well as in the porous matrix in between, defining $\Omega_M$. This model allows for integrating fractures of a length much smaller than the size of the domain into $\Omega_M$, providing secondary permeability, rather than as part of the explicitly characterized fracture network.  The DFM model is appropriate for handling situations when a permeable medium hosts fractures that should be explicitly represented due to their structural and dominating impact of processes.  

In principle, all fractures within the domain can be incorporated into $\Omega_F$, and the governing equations are then given in Sect. \ref{sec: Modeling of flow in fractures and matrix}.
In practice, it is not feasible to retain an explicit representation of all fractures, mainly for reasons of computational complexity, as discussed in Sect. \ref{sec:discretization}. DFM models therefore preserve some fractures in $\Omega_F$, while others are upscaled, and replaced by averaged quantities within $\Omega_M$. Hence, the effective equations will be those of a fractured porous medium, even if the host medium is considered impermeable.
By resolving the interface between fractures and the rock, DFM models facilitate detailed modeling of processes on this interface, including capillary pressures in two-phase flow, and hydro-mechanical interaction between flow and the host medium. 

Methods to decide which fractures to assign to $\Omega_F$ and $\Omega_M$, respectively are not well developed; common practice is to base the selection on fracture length \citep{Lee2001}.
More advanced, and application dependent, selection criteria could include factors such as fracture connectivity and the volume of host material that has a fracture as its main connection to the global flow field could be included.
The type of upscaled representation can also be varied: The traditional approach has been a single-continuum method \citep{Karimi-fard2004}, but multi-continuum approaches can also be applied \citep{Sandve2013, Jiang2017}. 

\subsubsection{Discrete fracture network models}
The network of fractures, disregarding the host medium, is commonly referred to as a discrete fracture network (DFN). By analogy, models that ignore flow in $\Omega_M$, or equivalently considers $\Omega_M$ impermeable,  are referred to as DFN models. 

In a DFN model, all fluid is assumed to be contained within the fracture network, which is represented by a lattice. As the material in between the fractures is considered impermeable, this is strictly speaking not a model of a fractured porous media, but we still include the description of the model for completeness. In particular DFN flow models are appropriate model of porous media where the entire porosity and permeability is due to fractures that can be explicitly represented. The model is commonly also used as a generalization to model fractures in low-permeable porous media. 

In DFN models, the fracture network is represented as a lattice. Flow and transport are modeled in individual fractures $\Omega_F^i$ based on Eq. \ref{eq:Darcy_f}, with network interaction via the intersections $\Omega_I$, as discussed in Sect. \ref{sec: Modeling of flow in fractures and matrix}.
This gives an accurate representation of dynamics in fractures, which can be of use in multi-physics couplings, such as reactive flows and and mechanical deformation of fractures.

\section{Discretization approaches}
\label{sec:discretization}
The conceptual models presented in the previous section can all be used as starting points for simulation models, and this section provides an overview of prevailing techniques and challenges.
The main topics are the preservation of heterogeneities between fracture and host medium, and  discretization of the interaction between the two.
The full simulation model will also require discretization of the flow within the fractures and, in most cases, in the matrix. We will comment on this when appropriate, but generally refer to \citet{Lie2016}  for an introduction to discretization of porous media flow.
Quantitative comparisons between a number of different numerical approaches are presented for DFM models in \citet{Flemisch2018}, DFN models in \citet{Fumagalli2018} and DFM and dual-continuum models in \citet{Moinfar2013}.

\begin{figure}
\centering
	\includegraphics[width=1\textwidth]{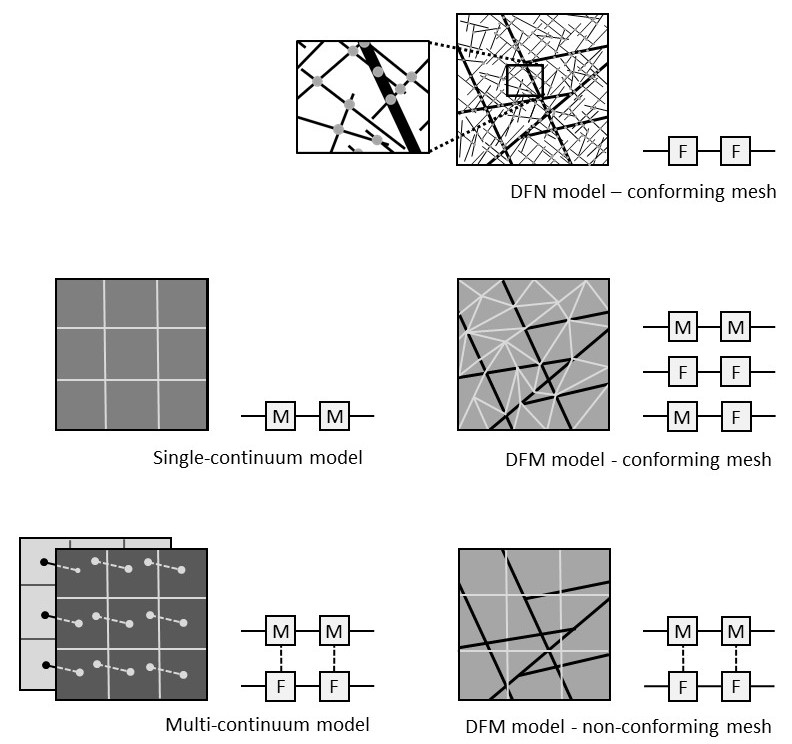}
	\caption{Illustration of discretization approaches. For each approach an conceptual illustration of the mesh is given to the left, while the types of cell couplings resulting from the model is shown to the right. Vertically stacked boxes linked with dashed lines denote degrees of freedom at the same physical point. The approaches based on implicit representation of fractures are shown to the left, while the approaches based on explicit representation of fractures are shown at the right. For the multi-continuum models we show a dual medium only.}
	\label{fig:discretization_couplings}       
\end{figure}

Closely related to the conceptual models discussed in Sect. \ref{sec:conceptual models} which treat fractures implicitly (single- and multi-continuum models) or explicitly (DFN an DFM models), we divide the numerical approaches into three groups: (i) models with implicit fracture representation (ii) models with explicit fracture representation with non-conforming meshes to the fractures, and (iii) models with explicit fracture representations with conforming meshes to the fractures. The different variants are illustrated in Fig. \ref{fig:discretization_couplings}.

\subsection{Implicit fracture representation}
When fractures are represented implicitly, the model contain no direct information of fracture geometry, and are, hence, discretized with meshes that are non-conforming to the fractures.
\subsubsection{Single-continuum approaches}
Single-continuum models are by far the simplest approaches with respect to discretization: At any point in space there is a single value for each parameter such as permeability, porosity.
This leaves us with a standard simulation model of porous media flow, that can be dealt with by standard approaches for simulation of flow in porous media.

\subsubsection{Multi-continuum approaches}
Depending on which type of multi-continuum approach taken, these models will have several coexisting values for physical parameters, each assigned to a separate medium. As for the single-continuum models, discretization of the individual media can be treated as standard (non-fractured) porous media flow, applying any number of numerical techniques. The main difference thus lies in the treatment of the interaction term: For linear approximations, this will not significantly alter the characteristics of the system, although the stiffness of the problem is increased and may cause convergence issues for linear and non-linear solvers. More advanced interaction terms (non-linear, non-local or time dependent) may substantially increase the complexity of the problem. For each included medium  the number of degrees of freedom will increase significantly (number of physical cells for single-phase flow, twice the number for two-component flow etc). This can substantially increase the computational cost of the simulation, in particular due to increased time spent by linear and non-linear solvers.

\subsection{Explicit fracture representation with non-conforming meshes}

The above approaches have in common that upscaling removed the distinction between individual fractures prior to the introduction of a numerical mesh. 
If the conceptual model takes a conforming approach to representing the  fractures, these must either be explicitly resolved by the mesh, as discussed  in the next section, or upscaled as part of the discretization as discussed  here.
The resulting simulation models typically have separate discrete variables for fractures and matrix. Difficulties that needs to be addressed are associated with i) the amount of geometric complexity contained within individual cells, ii) discontinuities in matrix pressure across the fracture, caused by resistance to flow according to Eq. (\ref{eq:interface_flux}).

\subsubsection{Embedded discrete fracture methods}
Embedded discrete fracture methods (EDFM), introduced by \citet{Li2008}, are based on the DFM conceptual model, and discretized with separate degrees of freedom for $\Omega_M$ and $\Omega_F$. Flow within $\Omega_M$ and $\Omega_F$ is discretized separately, the latter commonly borrows techniques for conforming discretizations of DFM models discussed in Sect. \ref{sec:dfm_conf_disc},  see e.g.  \citet{Moinfar2014}.
Flow between fracture and matrix within a matrix cell is taken as proportional to the pressure difference between fracture and matrix, with a proportionality constant reflecting the geometry of the fracture.
The coupling structure of the discretized system is identical to that of dual-continuum models, see Fig. \ref{fig:discretization_couplings}, with the difference that the coupling terms is modeled in terms of discrete variables directly.

\subsubsection{Extended finite element methods}
While the methods discussed so far can be implemented using several 
discretization schemes, the extended finite element method (XFEM) is tightly 
bound to finite element methods.
The method was originally developed for fracture mechanics, but has been 
applied to flow problems, see e.g. \citet{Fumagalli2013,Schwenck2015,Flemisch2016}.
For non-conforming meshes, a finite element approximation to the pressure will be poor in cells containing a fracture, as the finite element basis functions cannot capture the discontinuity of pressures over the fracture surface.
To improve on this, XFEM enriches the finite element space by adding basis functions designed to capture the discontinuity.
While the idea is simple, the number of, and design of, the basis functions depends on the geometric configuration of the fractures within the cells, and implementation becomes difficult for fracture crossings, endings within a cell etc, see \citet{Schwenck2015} for details. These complications increase further in 3D, and to date, XFEM for flow problems has been used almost exclusively in 2D.


\subsection{Explicit fracture representation - conforming mesh}
\label{sec:conf_disc}
Conforming methods avoid the computation of transfer functions  $\Omega_M$ and $\Omega_F$  by explicitly meshing  $\Omega_F$ and $\Omega_I$. 
This gives high resolution of near-fracture dynamics, which can be especially useful for multi-physics couplings.  
The price to be paid is in mesh construction, which can be extremely difficult, and computational cost, as the simulation model will now in general contain many more cells.

\subsubsection{Meshing}
\label{sec:meshing}
A conforming mesh must by definition explicitly resolve $\Gamma_{MF}$ by treating fractures as constraints in the mesh construction. By extension, the mesh should also conform to $\Omega_I$.
Unless the fracture geometry is highly regular, fitting a structured mesh with a Cartesian topology to the fracture surfaces is virtually impossible, and unstructured meshing algorithms mush be applied.
Most algorithms apply simplex meshes (triangles in 2D, tetrahedrals in 3D), but Voronoi meshes have also been considered \citep{Merland2014}

\begin{figure}
\centering
\begin{subfigure}[Parallel]{
\includegraphics[height=0.2\textheight]{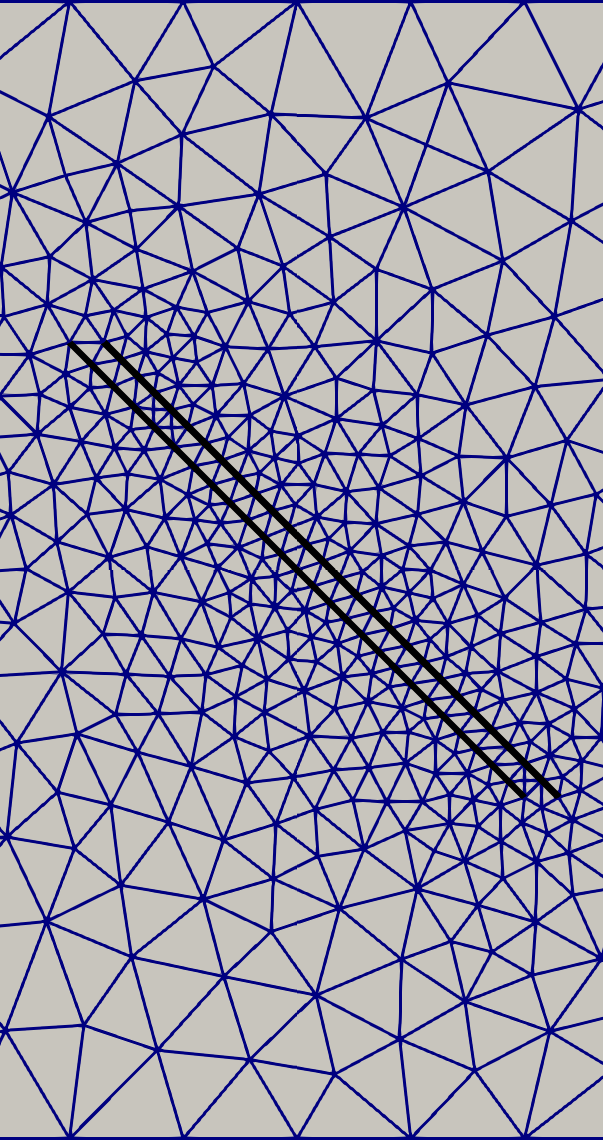}
}
\end{subfigure}
\begin{subfigure}[Small angle]{
\includegraphics[height=0.2\textheight]{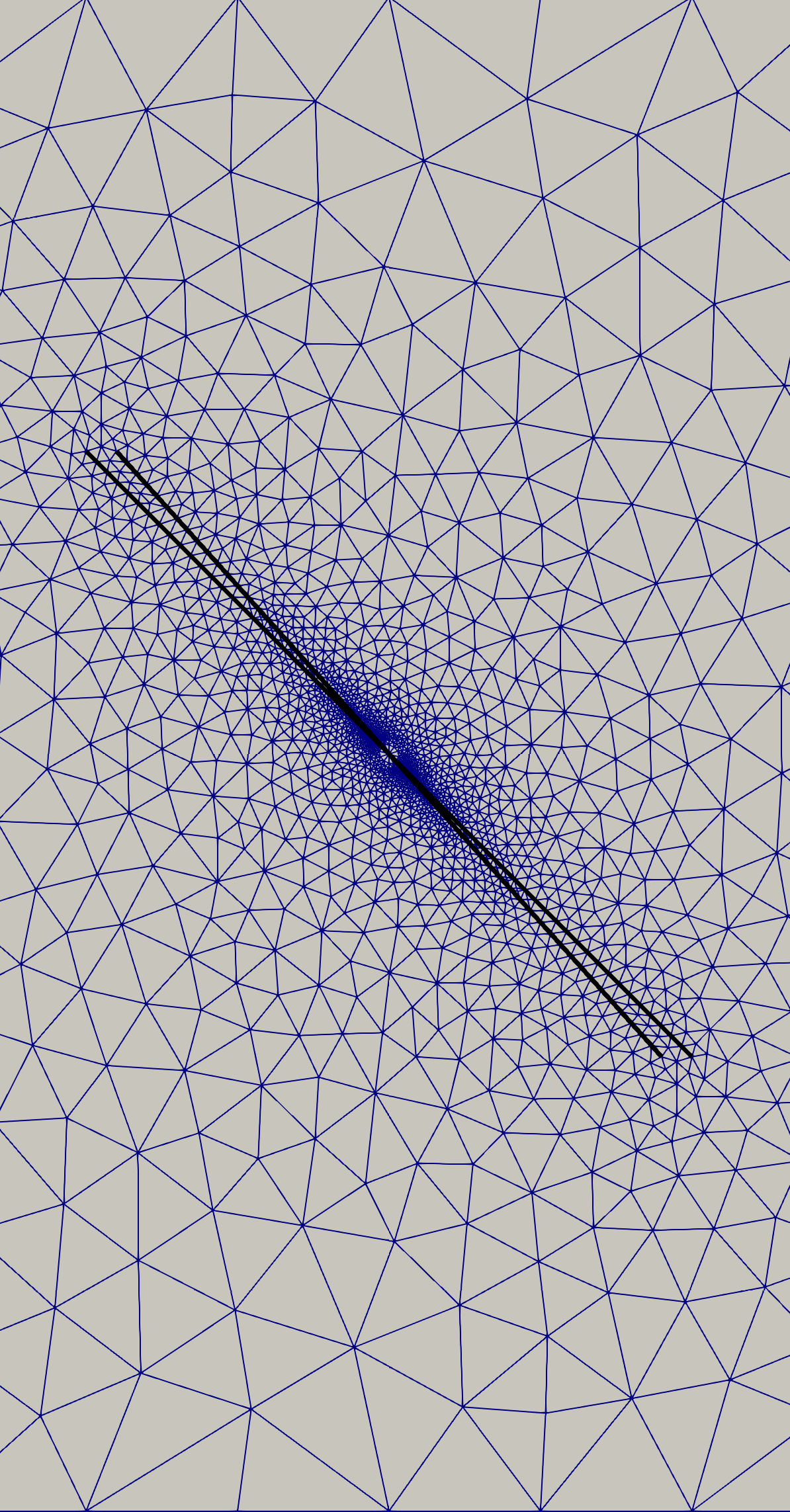}
}
\end{subfigure}
\begin{subfigure}[X-intersection]{
\includegraphics[height=0.2\textheight]{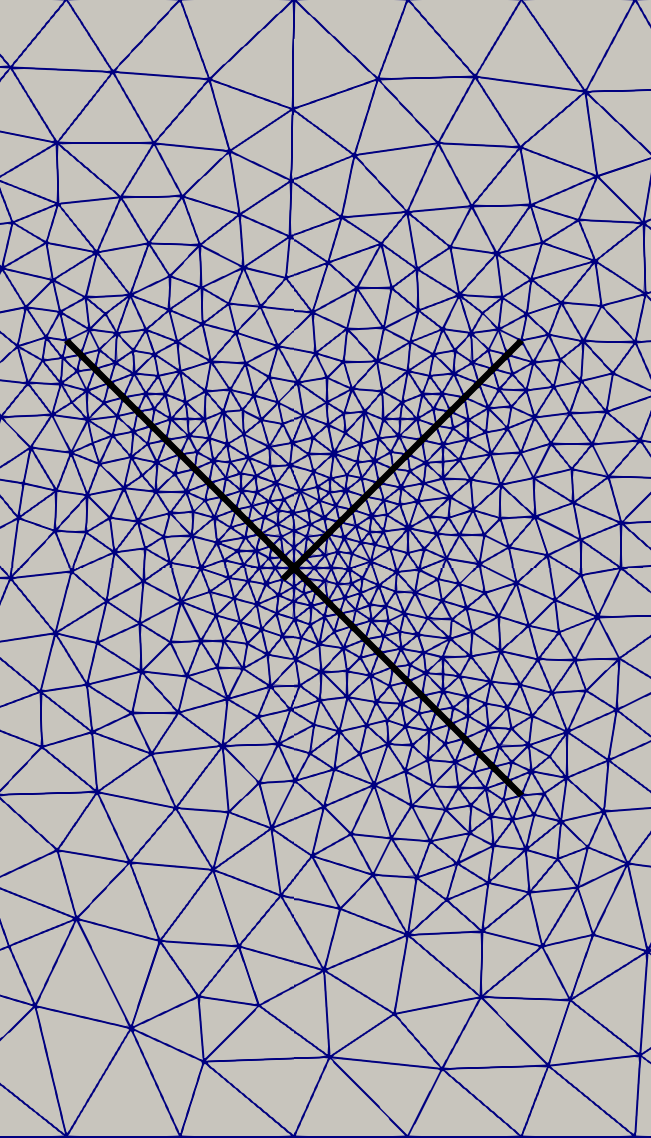}
}
\end{subfigure}
\begin{subfigure}[Almost crossing]{
\includegraphics[height=0.2\textheight]{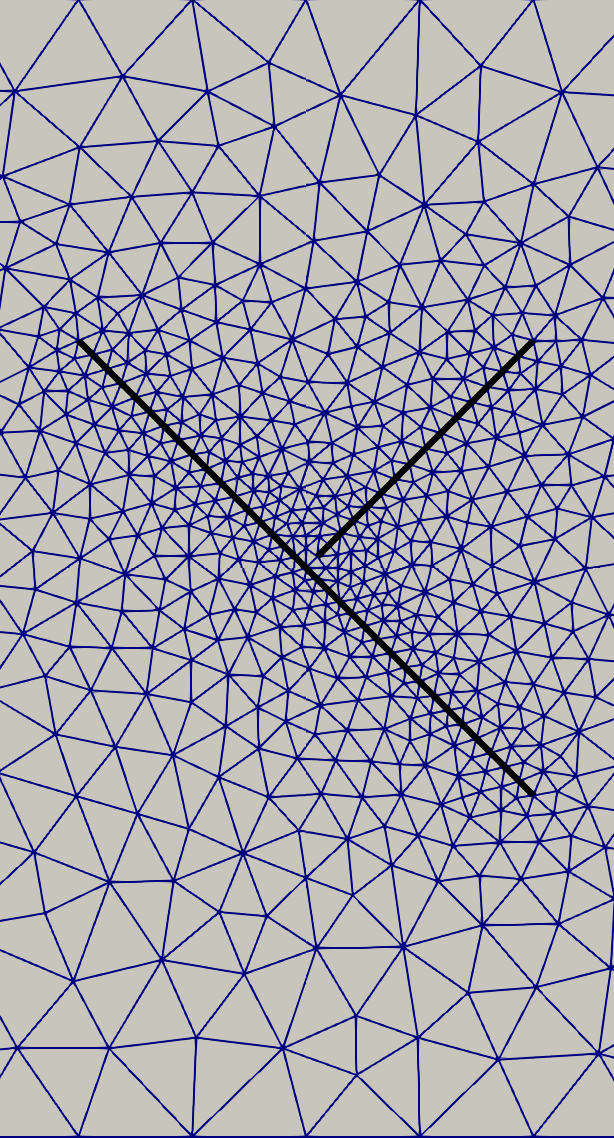}
}
\end{subfigure}
\caption{Illustration of fracture configurations that pose difficulties for conforming meshing algorithms.}
\label{fig:meshing}
\end{figure}

To understand why simulations on conforming meshes can be challenging, recall that the performance of numerical methods in general is tied to the shape and number of of  cells: 
Most method will perform poorly on cells with small angles, often yielding low accuracy solutions for extreme cases.
Moreover, computational cost in the best case increase linearly with the number of cells in the mesh, and the effectiveness of linear and non-linear solvers will often depend on the ratio of the smallest to the largest cell, depending on whether effective preconditioners are available.
In addition, the smallest cell size is bounded above by the size of the smallest fracture to be explicitly represented, and in practice the cells will often be much smaller.

In the case of an equi-dimensional representation of the fractures, a very fine mesh is needed to resolve the fractures and at the same time avoid large aspect rations of the fracture cells. With the mixed-dimensional representation of the fractured media that was introduced in Sec. \ref{sec:mixed-dimensional} the dimension of the fracture cells are one dimension lower than that of the matrix cells, and this issue is avoided. Nevertheless, meshing remains challenging for complex network geometries, as discussed below.

Figure \ref{fig:meshing} shows four different configurations known to cause problems in 2D. 
All of these configurations will either require highly refined meshes or cells with high aspect ratios; the fractures in Fig. \ref{fig:meshing}b) will also necessarily require cells with sharp angles.
The complications are even more severe in 3D. 



Meshing algorithms used in simulations can either be general purpose, drawing upon sophisticated software packages such as \citet{Geuzaine2009,Si2015,Shewchuk1996}, or dedicated to fractured porous media \citep{Holm2006}.
Interestingly, the latter class includes approaches that allows slight modifications in the fracture geometry to ease the mesh construction \citep{Karimi-Fard2016,Mallison2010}, or even replacement of troublesome fractures in a stochastic setting \citep{Hyman2014}.
Even when using such adaptations, meshing remains a bottleneck for conforming discretizations due to the inherent geometric complexity of fracture networks.

\subsubsection{Discretization approaches}
\label{sec:dfm_conf_disc}
Discretization of conformal models can be categorized as equi-dimensional, mixed-dimensional and hybrid approaches.
The model for the equi-dimensional approach is essentially that of a standard porous medium, with heterogeneous parameters in $\Omega_M$ and $\Omega_F$, and we will not discuss this further.
Discretization of mixed-dimensional conforming methods is naturally treated as a domain decomposition method, wherein the flow equations are discretized on $\Omega_M$, $\Omega_F$ and $\Omega_I$ separately, leaving out $\Omega_M$ in DFN models, 
and then coupled by interface conditions on $\Gamma_{MF}$ and $\Gamma_{FI}$.  For the flow problem, this approach can be shown to be stable under very mild assumptions on the discretizations 
\citep{Martin2005,Angot2009,Boon2016,Nordbotten2018}. 
The hybrid approach to conforming discretizations is based on equi-dimensional modeling. However, in mesh construction fractures are treated as lower-dimensional objects (surfaces), thus avoiding difficulties of meshing close parallel surfaces. In the discretization of flow equations, faces on the fracture surfaces are then expanded into equi-dimensional cells by a virtual extension perpendicular to the fracture surface. This is a convenient way to introduce a fracture model in an existing simulation model, see e.g. \citet{Karimi-fard2004}.

The computational cost of a conforming method can in part be traced to cells in $\Omega_I$, which typically have much smaller volume than cells in $\Omega_M$ and $\Omega_F$. The results in badly conditioned linear systems in the flow problem, and time stepping constraints for explicit transport schemes.  This can be circumvented by eliminating variables in $\Omega_I$  \citep{Karimi-fard2004,Sandve2012}, with a varying loss of accuracy \citep{Stefansson2017}.



\section{Discussion}
\label{sec:discussion}
Over the last decades, modeling and simulation of single-phase flow in fractured porous media has matured as a research field. Different approaches and methods presented in the previous sections handle a range of scenarios based on the underlying structure of the fracture network. 

In this section, we first discuss aspects that may enter in selecting a specific model, before discussing extensions to multi-phase flow and more general multiphysics couplings.

\subsection{Characteristics of different modeling and discretization approaches}
A main remaining challenge in modeling and discretization of flow in fractured porous media is the appropriate selection of model and discretization approaches. 
This selection is closely linked to the specific modeling problems. In some cases, a continuum model, where the fractures are integrated in an effective porous medium, will be sufficient. This is primarily when fractures are weakly connected and small compared to the characteristic length scales of interest in modeling. If fractures occur as dense connected networks, multi-continuum models can capture the different characteristic fluid velocities in the fractures and the surrounding domains. When the local geometry of the fracture network affect directly the flow patterns at the scale of interest, fractures must be represented directly, by DFM or DFN models. 
For these models, we have seen that a lower-dimensional representation of fractures is beneficial from a modeling and, in particular, a meshing point of view.
A combination of models, by restricting different models to different parts of a domain, or by using models with explicit representation of fractures as basis for upscaled continuum models, is also an alternative. This implies a balancing between discretization errors, directly associated with the mesh size and quality, and modeling errors caused by upscaling fractures. Little is known of how these errors relate. Furthermore, modeling choices are related to application-specific  availability and quality of data, as well as modeling purposes.  


When aiming for simpler effective representations of fractured porous media, we also have to account for the challenge that fractures introduce new scales compared to the standard micro-scale (pore) and macroscale (continuum scale) models of porous media. This results in a lack of scale separation, and a proper REV may not exist. Depending on how pronounced these features are, fractured porous media are more or less suited for upscaled representations. However, constraints on resources, as well as available information, often trigger a need for at least a partial effective representation of dynamics, often relying on engineering and physical intuition to balance errors and costs. 



In general, we have seen that the different options for conceptual model and discretization move the complexity, and the error,  between modeling and discretization.
This is best illustrated by comparing a non-conforming approach, be it multi-continuum or EDFM, and a conforming approach. From a discretization standpoint, non-conforming methods do not require complex meshes, and simulations are fast and simple compared to conforming methods. The price to be paid is in complex modeling and low resolution of fracture-matrix interaction.
Fundamentally, these issues are caused by the inherent complexity of porous media flow dominated by fracture networks, and modeling of the flow will remain difficult.

\subsection{Extensions to multi-phase flow}
The focus of the previous sections has been the modeling of single-phase flow in fractured media. However, in many applications two or more fluid phases share the pore-space. The interplay of capillary forces, pore geometry, buoyancy and viscous forces then leads to non-linearly coupled flow. Despite more than a century of research, the modeling of these phenomena across different length scales still forms an active research field and many challenges translate directly to fractured media. We do not give a detailed review but rather provide a brief discussion on challenges and activities. We group the discussion into three themes: 
(i) the conceptual and constitutive representation of multi-phase flow phenomena in a single fracture, (ii) the mathematical and numerical representation of conventional multi-phase Darcy terms in explicit representations of fractures and (iii) the representation of multi-phase flow phenomena in implicit representation of fractures. 

The availability of experimental or numerical data for the constitutive relationships, capillary pressure and relative permeabilities, is surprisingly sparse compared to non-fractured media. Since the first experimental works by \citet{Pruess1990,Persoff1995} and modeling work by \citet{Rossen1992} progress has been achieved \citep{Bogdanov2003, Chen2006}. However, characterization of flow patterns in the fracture is still an active research field  \cite{Karpyn2007,Babadagli2015, Jones2017}, as are which impact they have on constitutive functions and their upscaling.

Assuming constitutive relations for relative permeabilities and capillary pressures are known, they can in principle be directly included in models with explicit fracture representations. However, differences in the composition of fracture and matrix may cause strong heterogeneities in the constitutive relations, manifested as capillary barriers and entry pressure effects, over the fracture-matrix interface. 
Depending on the specific form of relative permeabilities and capillary pressure functions, these heterogeneities are known to cause numerical challenges, see e.g. \citet{Jaffre2011,Ahmed2017,Brenner2018}. 

The representation of multi-phase phenomena in continuum formulations is more challenging. As the transfer time-scales are much slower than in single-phase flow, single-continuum representations are less likely to work because the local equilibrium assumption that underlies conventional relative permeabilities and capillary pressure relationships is not given. 
A numerical upscaling approach for parameters has however been outlined by \citet{Matthai2009}.  The single-phase multi-continuum formulations have been expanded to incorporate multi-phase flow from early on \citep{Kazemi1976, Gilman1988} and continue to be developed \cite{Lu2008,Tecklenburg2013,March2018}. Comprehensive reviews are given in \citep{Lemonnier2010a,Lemonnier2010b} and \citep{Ramirez2009a, Ramirez2009b}. The fundamental challenge is that for a systematic development one would want to isolate different processes and develop transfer concepts individually while these processes are non-linearly coupled. 

\subsection{Outlook - multiphysics couplings}
This paper has reviewed the main modeling and discretization approaches for flow in porous media. However, what is often characteristic of fractured porous media, is the strong mutual interaction between flow processes and the fractured structure itself.  Flow processes can lead to deformation of the fractures due to mechanical, thermal and/or chemical processes, and altered fracture configurations again strongly impacts flow.
This is for instance the case considering membranes or thin layers in living tissues \cite{Vernerey2012}. Another example is flow-induced seismicity, where flow reactivates and open fractures that again severely affects flow patterns \cite{Ucar2017}.

The coupled thermal-hydraulic-mechanical-chemical (THMC) processes in the context of fractured porous media pose research challenges that have only partially been solved, despite strong research contributions over the last two decades \cite{rutqvist2002modeling, taron2009numerical,kolditz2016thermo}.
For future research on flow in fractured media, inherent and remaining research 
challenges related to conceptual modeling and discretization approaches 
can be summarized in the following research questions
\begin{itemize}
\item How can disparate processes with different characteristic length and timescales be coupled in the mathematical models and discretization approaches presented herein? 
\item How can macroscale representations of fractured porous media be amended to include not only changing fracture apertures, but also propagating fractures?
\item In modeling of dissolution and precipitation processes, how should decisive phenomena such as clogging of fractures be modeled?
\item What are suitable upscaled models for interacting processes in fractured media?
\item How can we use modeling for qualitative and quantitative insight when data on fracture configurations and information on constitutive relationships are poor?
\end{itemize}

\section{Conclusion}
\label{sec:conclusion}
This paper has reviewed mathematical and physical models 
and numerical approaches developed over the last decades to model flow in fractured porous media. While this is now a mature research field, providing modeling concepts and discretization approaches to handle a range of situations where networks of fractures dominate flow, 
substantial challenges remains in dealing with multi-phase flow and multi-physics couplings (multi-phase flow, mechanics, chemistry).
\section*{Acknowledgements}
This work was partly funded by the Research Council of Norway through grants no. 267908/E20, 244129 and 250223.

\bibliographystyle{plainnat}      
\bibliography{Inga_refs,refsEK,refFD}   
\end{document}